\documentclass[showpacs,pre,aps,amssymb,amsmath]{revtex4}
\usepackage{graphicx}
\newcommand{\be}{\begin{equation}}
\newcommand{\ee}{\end{equation}}
\newcommand{\bd}{\begin{displaystyle}}
\newcommand{\ed}{\end{displaystyle}}

\begin{document}
\title{Space-Time Complexity in Hamiltonian Dynamics}


\author{V. Afraimovich}

\affiliation{San Luis Potosi University - IICO, Av. Karakorum 1470,
San Luis Potosi, SLP-78240, Mexico}

\author{G.M. Zaslavsky}

\affiliation{Courant Institute of Mathematical Sciences,
New York University,
251 Mercer St., New York, NY 10012, and}

\affiliation{Department of Physics, New York University,
2-4 Washington Place, New York, NY 10003, USA }

\date{\today}

\begin{abstract}

New notions of the complexity function $C(\epsilon ;t,s)$ and entropy
function $S(\epsilon ;t,s)$ are introduced to describe  systems
with nonzero or zero Lyapunov exponents or systems that exhibit strong
intermittent behavior with ``flights'', trappings, weak mixing, etc.
The important part of the new notions is the first appearance of
$\epsilon$-separation of initially close trajectories. The complexity
function is similar to the propagator
$p(t_0 ,x_0 ;t,x)$ with a replacement of $x$ by the natural
lengths $s$  of
trajectories, and its introduction does not assume  of the
space-time independence
in the process of evolution of the system. A special stress
is done on the choice of variables and the replacement
$t\to \eta = \ln t$, $s \to \xi = \ln s$ makes it possible to consider
time-algebraic and space-algebraic complexity and some mixed cases. It
is shown that for 
typical cases the entropy function $S(\epsilon ;\xi ,\eta )$ possesses
invariants $(\alpha ,\beta )$ that describe the fractal dimensions of the
space-time structures of trajectories. The invariants $(\alpha ,\beta )$
can be linked to the transport properties of the system, from one side,
and to the Riemann invariants for simple waves, from the other side. This
analog provides a new meaning for the transport exponent $\mu$ that can be
considered as the speed of a Riemann wave in the log-phase space of the
log-space-time variables.
Some other applications of new notions are considered and numerical
examples are presented.

\end{abstract}

\maketitle

\section*{Lead Paragraph}

It is found in many cases that Hamiltonian chaotic dynamics possesses
in many cases a kinetics that doesn't obey the Gaussian law process and
that fluctuations of the observables can be persistent, i.e. there is no
any characteristing time of the fluctuations decay. This type of
dynamics can be characterized by the so called polynomial complexity
rather than an exponential one. More accurately, one can introduce some
complexity function and entropy function based on the dynamical process
of separation of trajectories in phase space by a finite distance during
a finite time. The new approach to the problem of complexity and entropy
covers different limit cases, exponential and polynomial, depending on
the local instability of trajectories and the way of the trajectories
dispersion.

\section{Introduction}

\label{S:intro}

The complexity of dynamical systems begins from the description of
chaotic trajectories in phase space
(\cite{he,ti}).
The notion of complexity has a rigorous meaning and it presents a quantity
that characterizes systems and can be measured. In an oversimplified way
one can say that the less predictable is a system, the larger complexity
should be assigned to the system. The original version of the dynamics
complexity was closely linked to the system's instability and entropy
\cite{KT,b,ti}.
A typical situation of the chaotic dynamics could be associated with a
positive Kolmogorov-Sinai entropy, or be similar to the Anosov-type systems.
This type of randomness and complexity of the systems can be
characterized by the exponential divergence of trajectories in phase space.

As long as investigation of chaotic dynamics reveals new
and more detailed pictures of chaos, the simplified old version of the
complexity appears to be constrained to be applied to typical systems.
Let us mention that the typical Hamiltonians do not possess ergodicity,
the boundary of islands in phase space make the dynamics singular in their
vicinity, and even zero measure phase space domains in the Sinai billiard
are responsible for the anomalous kinetics
\cite{za1}.
Attempts to find  adequate complexity definitions for realistic chaotic
dynamics were subject to many publications and reviews
\cite{p,gp,abc,bp}.
The basic idea of new developments for chaotic systems is to involve a
finite time of the systems unstable evolution into a definition of the
complexity or entropy. In some sense, our paper is a continuation of these
attempts.

There are numerous observations that the Hamiltonian systems referred as
the chaotic ones, {\it do not have} exponential dispersion of trajectories
for arbitrary long time intervals. These pieces of trajectories, called
flights, appear with a probability that is not exponentially small (see
for example
\cite{zen,bkwz,za2}).
A similar type of random dynamics with zero Lyapunov exponent appears in
billiards and maps with discontinuities
(\cite{ch,lv,ze1,ze2,aft}).
The behavior of systems with zero Lyapunov exponents definitely have some level
of complexity and some value of entropy in a physical sense, but the same
systems with zero Lyapunov exponent cannot be applied to by the regular
notion of the Kolmogorov-Sinai entropy of the standard definitions of
complexity. The most appropriate thing to say about such systems is that the
proliferation of an indefiniteness has an algebraic dependence on time
rather than the exponential one. Moreover, some systems behave in a mixed
way: partly with an exponential growth of their enveloping
(coarse-grained) phases volume and partly with its algebraic growth
with time.

All the above mentioned facts lead us to a necessity to introduce a new
notion of complexity and entropy and this is the goal of this paper. Our
scheme is based on a similar one to the Bowen idea of the
$\epsilon$-separation of trajectories, which explicitly imposes the
instability features of the dynamics. However, instead of the complexity
$C(\epsilon ,t)$ we introduce a complexity function $C(\epsilon ;t,s)$
which is similar to a propagator $p(t_0 ,x_0 ;t,x)$ with a replacement
$(x_0 ,x) \to (0,s)$ where $s$ is the natural length of trajectories
between $(x_0 ,x)$. The corresponding entropy $S(\epsilon ;t,s)$ can be
defined as $\ln C(\epsilon ;t,s)$. In this form the complexity and
entropy describe the system evolution in space-time without assumption of
the space-time separation.

Another important change is related to the choice of {\it basic variables}
which can be $\xi = \ln s$, $\eta = \ln t$ instead of $(s,t)$.
Similar logarithmic variable $\ln x$ instead of $x$ appeared in
\cite{mosl}
for a definition of entropy when L\'{e}vy distributions is considered.
We show
that for some typical cases $C(\epsilon ;\xi ,\eta )$ possesses invariants
$(\alpha ,\beta )$ similar to the Riemann invariants for the simple wave
propagations. The invariants $(\alpha ,\beta )$ do not depend on $\epsilon$
and they represent space-time fractal dimensions of the dynamical system. In
the proposed way of description of random dynamics, the notions of
{\it complexity function} and {\it entropy function} appear to be
constructive tools of the description of dynamics with zero or nonzero
Lyapunov exponents, with mixing or weak mixing properties, and with normal
or anomalous transport. A similar approach can be developed for a system
with mixed features when the basic variables are $(t,\xi )$ or $(\eta ,x)$.
In all considered cases the {\it entropy is an additive function} of the
corresponding basic variables.

As an important consequence of the new notions, we consider directional
complexity and directional entropy generalizing the notion introduction
in
\cite{m1} and \cite{acfm}.
\vspace{.2in}

\section{Preliminary comments on the complexity in phase space}
\label{sec:sec2}

Consider a dynamical system on the 2$n$-dimensional torus
${\bf T}^{2n}$ and let the dynamics of a particle be defined by an
evolution operator
\begin{equation}
\hat{T} (t) : \ (p(t), q(t)) = \hat{T} (p(0), q(0))] ,
\label{eq:eq2.1}
\end{equation}
which preserves the measure $\Gamma (p,q)$ = const (phase volume), and
$p,q \in {\bf R}^n$ are generalized momentum and coordinate. What kind of
dynamics should be considered as a simple one and what as a complex one?
We do not assume that there is the only definition of complexity which
particularly depends on how a notion of it will be applied to the dynamics.
A more or less typical definition of the complexity depends on how
trajectories are mixed in phase space due to the dynamics (\ref{eq:eq2.1}).
The stronger is mixing, the more complex is the dynamics. One can
immediately comment on some weak features of this type of approach which
deals with global phase space and global mixing. The process of mixing can
be non-uniform in time and it can have local space-different rates. We
call these features of mixing as space-time non-uniformities and, speaking
about the space, we have in mind the full phase space or its part where
the dynamics is ergodic.

The former comment leads us to a possibility of such definition of the
complexity which could embrace non-uniformity of space-time dynamic
processes represented by trajectories.

\begin{figure}[h]
\includegraphics[width=0.5\textwidth]{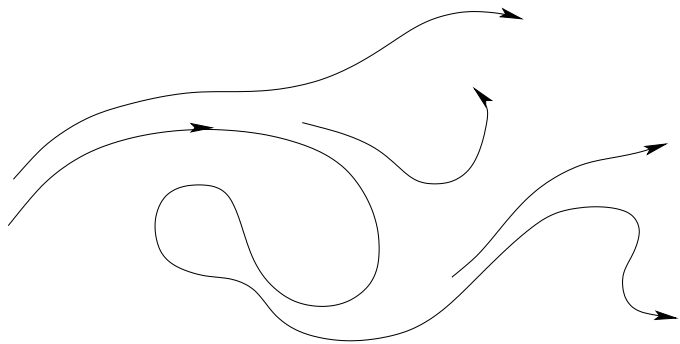}
\caption{Dispersion of trajectories in phase space.}
\label{fig1}
\end{figure}

Space-time non-uniformity suggests that vicinities of any trajectory
considered at different points $(p_j ,q_j ; t_j )$ with $p_j = p(t_j )$,
$q_j = q(t_j )$ may have very different dynamics of trajectories
(see FIG.~\ref{fig1}).
A strong ``inconvenience'' of this conclusion becomes clear if we assume the
``vicinity'' as an infinitesimal ball of the radius $\epsilon\rightarrow 0$
around a point $(p_j ,q_j ;t_j )$: it is difficult, if not impossible, to
describe a trajectory finite-term behavior on the basis of the
information about the trajectory from an infinitesimal domain of phase
space. The necessary data should arrive from finite pieces of trajectories
which make a possible definition of complexity to be space-time non-local.
We should be ready to have a situation with an exponential divergence of
trajectories at some small parts of the phase space, and to have a
sub-exponential divergence at other parts.

The importance of observing a trajectory with some precision $\epsilon$
during a finite time $t$ was discussed in detail by Grassberger and
Procaccia in [GP] and [P]. They also consider fluctuations of the Lyapunov
exponents obtained from a finite time observation. These works also
presented a space-time partitioning as a way to obtain correlation
properties of trajectories. Our analysis here will be extended, comparing
to [GP] and [P], in two directions: it will be applied to dynamical systems
that may have a sub-exponential divergence of trajectories at some phase
space domains, and it will deal with area-preserving Hamiltonian dynamics,
which  permits the use of some important results on the Poincar\'{e}
recurrences.

\section{Definitions of complexity in dynamical systems}
\label{sec:sec3}

Here we present a brief review of a few important definitions of complexity
of dynamics that involve a bunch of orbits and their comparable behavior
in phase space.

\subsection{Symbolic complexity}
\label{sec:sec3.1}

Historically, the first notion of complexity was introduced by
Hedlund and Morse [He] for symbolic systems. Let
$\{ \omega\} = (j_0 ,j_1, \ldots ,j_N, \ldots )$
be an admissible sequence  and
$\Omega = \{ \omega \}$ be
the  set of all admissible sequences.
One can say that $\{ j_k \}$ are coordinates,
$\omega_n :=\{j_0 ,j_1 , \ldots,j_{n-1}\} $ is a piece of an
admissible trajectory of length $n$, and $\Omega$ is the phase space.
Dynamics is defined by a shift operator
\begin{equation}
\hat{T} (j_0 ,j_1 ,j_2 ,\ldots ) = 
(j_1 ,j_2 ,\ldots ) \label{eq:eq3.1}
\end{equation}
The complexity $C_n (\omega )$ of an individual orbit going through a
point $\omega = \omega (j_0 ,j_1 ,\ldots )$ is the number of different words
$(j_k ,j_{k+1} ,\ldots ,j_{k+n})$ of the length $n$ $(k\geq 0)$ in the
sequence $\omega$. If we take into account that each word
$(j_k ,\ldots ,j_{k+n})$ indicates a piece of the symbolic phase space,
i.e. an element of the phase space partition, then $C_n (\omega )$ is the
number of different cells in the phase space available by the orbit going
through the initial point $\omega$.

It was shown in [He] that if $C_n (\omega )\leq n$ then $\omega$ is
eventually periodic, i.e. $C_n (\omega ) \leq$ const and, for example,
there is no such $\omega$ that $C_n (\omega ) \sim n^{1/2}$. The
complexity $C_n (\omega )$ may grow exponentially with $n$
\begin{equation}
C_n (\omega ) \sim \exp (hn) \label{eq:eq3.2}
\end{equation}
and in this case $h$ is the topological entropy. There are examples of
\begin{equation}
C_n (\omega ) \sim n^{\gamma} , \ \ \ \ (\gamma > 1 )
\label{eq:eq3.3}
\end{equation}
i.e. of the sub-exponential complexity (see more in [Ti], [Fer], and
references therein).

\subsection{Topological complexity}
\label{sec:sec3.2}

A transition from the symbolic complexity to the complexity of dynamical
systems with arbitrary topological phase space was suggested in [Bla].
Let a topological space $X$ and a continuous map $f^t : \ X \rightarrow X$
\begin{equation}
f^t x_0 = x_t \ , \ \ \ \ (x_k \in X)
\label{eq:eq3.4}
\end{equation}
generate a dynamical system $(\hat{T};X)$, and let
${\cal D} = ({\cal D}_1 ,{\cal D}_2 ,\ldots ,{\cal D}_n )$
is a finite cover of the phase space by, say open,
subsets. For any initial point $x_0 \in {\cal D}_{i_0}$ one can consider
different itineraries
$(x_0 ,\ldots ,x_k = f^k x_0 ,\ldots ,x_n )$
such that $x_k \in {\cal D}_{i_k}$. Then the topological complexity
$C({\cal D},t_n = n)$ is the number of different possible itineraries of the
temporal length $n$ for points $x_k$ in $X$. Evidently, this complexity
depends on the covering system $\cal D$.

If the system is chaotic, then
\begin{equation}
C({\cal D} ,n) \sim e^{hn}
\label{eq:eq3.5}
\end{equation}
where $h$ is the topological entropy provided that $\cal D$ is chosen in a
right way. Let us emphasize that the topological complexity deals with all
orbits of a dynamical system.

There are other possibilities to introduce a complexity such as Kolmogorov
complexity (see for example [Br]), a measure theoretical complexity [Fer1],
etc., which will not be discussed here.

\subsection{$\epsilon$-complexity}
\label{sec:sec3.3}

The definition of $(\epsilon,n)$-complexity  will be the most important
for our
following generalizations, and we discuss it in more detail.


In \cite{KT} the authors introduced notions of $\epsilon$-capacity and
$\epsilon$-entropy in space of curves which can or can not be solutions
of a differential equation. Let ${\cal L_T}=\{x(t), 0\leq t\leq T\}$, $x(t)\in M$
be a space of continuous curves endowed with the following (Chebyshev)
metric
\begin{equation}
d(\{x(t)\}, \{y(t)\}):= \max_{0\leq t\leq T} \rho( x(t),y(t))
\end{equation}
where $\rho$ is a metric in the space $M$. Let $N(\epsilon)$ be the
maximal number of curves which are $\epsilon$-pairwise $d$-disjoint.
Then $\log N(\epsilon)$ is called the $\epsilon$-capacity. Let $N_1(\epsilon)$
be the minimal number of sets of $d$-diameter $\geq 2\epsilon$, needed to
cover the set $\cal L_T$. Then $\log N_1(\epsilon)$ is called the
$\epsilon$-entropy.
If one assumes that $\{x(t)\}$ is a piece of an orbit of a dynamical system and
replaces the continuous time by discrete one, then one comes (as it was done
by Bowen [B]) to the following definitions.


Let us consider first a dynamical system generated by an evolution
operator $f$ on the phase space $M$. Let $A \subset M$ be a subset of
initial points (it could be invariant or not) and
$\ell_n (x) = \bigcup_{k=0}^{n-1} f^k x$ an orbit segment of temporal length
$n$ going through an initial point $x \in A$. Two segments $\ell_n (x)$
and $\ell_n (y)$, $x,y \in A$, are said to be $(\epsilon,n )$-separated
if there exists $k$, $0 \leq k \leq n-1$, such that
dist$(f^k x, f^k y) \geq \epsilon$, where dist means distance in the
phase space $M$. The maximal number of distinct segments of orbits with
accuracy $\epsilon$ is defined by
\begin{equation}
C_{\epsilon,n } (A) = \max \{ \# {\rm segments \ in \ a \ }
(\epsilon,n )-{\rm mutually \; separated \ set} \}
\label{eq:eq3.6}
\end{equation}
and is said to be
$(\epsilon,n )$-complexity of the set $A$. Evidently $\log C_{\epsilon,n }$ is
the $\epsilon$-capacity. It was shown in [B] (see also
\cite{D} for compact invariant subset of initial points) that
\begin{equation}
\label{eq:eq3.7}
h = h_{\rm top} (A) =:
\lim_{\epsilon\rightarrow 0} \
\overline{\lim}_{n\rightarrow\infty}
{\log C_{\epsilon,n} (A) \over n}
\end{equation}
is the topological entropy of the dynamical system $(f^k ,M)$ on the set
$A$, and in [T] that
\begin{equation}
\label{eq:eq3.8}
b =:
\lim_{n\rightarrow \infty} \
\overline{\lim}_{\epsilon\rightarrow 0}
{\log C_{\epsilon,n}(A) \over - \log\epsilon}
\end{equation}
is the upper box dimension of the set $A$. Thus, we may assume that if
$0 < b < \infty$, $0 < h < \infty$ then
\begin{equation}
\label{eq:eq3.9}
C_{\epsilon,n } (A) = \epsilon^{-b} \cdot e^{hn} \cdot \bar{C}(\epsilon ,n)
\end{equation}
where $\bar{C} (\epsilon ,n)$ is a subexponential function of $\ln\epsilon$
and $n$.

We will call the number $C_{\epsilon,n } (A)$ the Bowen or
$(\epsilon,n)$-complexity of the set $A$.

The definition (\ref{eq:eq3.6}) is fairly general and can be applied to
non-Hamiltonian and non-compact dynamics. The definition considers a set of
orbits without details of their separation process with a time-interval $n$
that typically is large.

To illustrate the property (\ref{eq:eq3.7}), consider one-dimensional
mixing dynamics on the interval
$[0 ,\ell ]$, $x \in [0,\ell ]$ with an exponential divergence of
trajectories. Let $\delta x_0$ is the initial distance at $t =0$ between
two trajectories and $\delta x_t  = \epsilon$ is the distance at time $t$.
Then
\begin{equation}
\label{eq:eq3.10}
\epsilon = \delta x_0 \exp ht
\end{equation}
i.e. for $\tau \geq t$  two trajectories are separated. The number of
such trajectories is
\begin{equation}
\label{eq:eq3.11}
C_{\epsilon ,t} (A) = {\ell_A \over \delta x_0} = {\ell_A \over \epsilon}
e^{ht}
\end{equation}
where $\ell_A$ is the length of a small initial interval $A$. The properties
(\ref{eq:eq3.6}) and (\ref{eq:eq3.7})
follow directly from (\ref{eq:eq3.11}) for $b = 1$.
If the set $A$ has the box-dimension $b$, then
(\ref{eq:eq3.10}) should be replaced by
\begin{equation}
\label{eq:eq3.12}
(\epsilon /\delta x_0 )^b = e^{ht}
\end{equation}
and correspondingly, instead of (\ref{eq:eq3.11})
\begin{equation}
\label{eq:eq3.13}
C_{\epsilon ,t} (A) = (\ell_A /\delta x_0 )^b = (\ell_A /\epsilon )^b
e^{ht}
\end{equation}
For more general situations we may assume
\begin{equation}
\label{eq:eq3.14}
C_{\epsilon ,t} (A) = (\ell_A /\epsilon )^b
e^{ht} \bar{C} (\epsilon ,t)
\end{equation}
where $\bar{C} (\epsilon ,t)$ is a
slow varying function of $\ln \epsilon$ and
$t$ compared to the main multipliers.

The expression (\ref{eq:eq3.14}) shows in an explicit way how the Bowen
complexity depends on the time interval $t$, accuracy $\epsilon$ to
determine the location of trajectories, and the domain $A$ of a set of
initial conditions. The dependence on $A$ can be eliminated from
(\ref{eq:eq3.11}) or (\ref{eq:eq3.13}) by choosing a normalized complexity
$C_{\epsilon ,t}$ per unit volume:
\begin{equation}
\label{eq:eq3.15}
C_{\epsilon ,t} = C_{\epsilon ,t} (A) /\ell_A^b ,
\end{equation}
which is possible due to the uniformity  of mixing in the considered
dynamical system.


{\bf Remark.} Complexity of an orbit.

 If one chooses an orbit
$\Gamma(x_0)= \cup_{k=0}^\infty f^k x_0$ in  the capacity of the set $A$
of initial points, one will arrive to a definition of the $\epsilon$-complexity
of the orbit $\Gamma(x_0)$. It is simple to see that this definition is
an analog of the symbolic complexity of a symbolic system described above.

It is not difficult to show that for any small $\delta>0$,
$$
C_{\epsilon(1+\delta),n} (\textup{clos}(\Gamma(x_0)))\leq C_{\epsilon,n}
(\Gamma(x_0))\leq C_{\epsilon,n}
(\textup{clos}(\Gamma(x_0)))\; ,
$$
i.e., the complexity of the closure of an orbit asymptotically behaves
in the same way as the complexity of the orbit.

Furthermore making use of the definition of complexity for an arbitrary
set $A$, one may introduce the complexity of a measure.

\noindent
{\bf Definition 1.} Given an invariant measure $\mu$, the quantity
\begin{equation}
C_{\epsilon ,t} (\mu):= \inf_A \{ C_{\epsilon ,t} (A) \; | \; \mu(A)=1\}
\end{equation}
ia called the complexity of the measure $\mu$.

We shall use this definition bellow.


\subsection{Complexity and  phase volume}
\label{sec:sec3.4}

Expression (\ref{eq:eq3.6}) can be interpreted in a way that may be generalized
to much more complicated situations. Let
again $M$ be the phase space, $\Gamma = \Gamma (M)$ its phase volume, and
$\Gamma_0$ be the  phase
volume
of a set of initial conditions $A_0 \subset M$
at time $t_0 = 0$ and consider
their evolution $A_t$ up to time $t$. Let $\bar{\Gamma}_t$ be a minimal
enveloping $A_t$ convex phase volume. Then for systems with
exponential divergence of trajectories
\begin{equation}
\label{eq:eq3.16}
\bar{\Gamma}_t = \Gamma_0 e^{ht} \ .
\end{equation}
To find how many different states can occupy the volume
$\bar{\Gamma}$, one should define an
``elementary'' minimal volume of one state, i.e. $\epsilon^b$.
Then
\begin{equation}
\label{eq:eq3.17}
\max_A \ \# \ {\rm states \ in} \ \bar{\Gamma}_t = (\Gamma_0 /
\epsilon^b )e^{ht}
\end{equation}
where maximum is considered with respect to different sets $A$ in
$\Gamma_0$.

Expressions (\ref{eq:eq3.11}),(\ref{eq:eq3.12})
permit an important physical interpretation.
Hamiltonian chaotic dynamics preserves the phase volume, i.e.
$\bar{\Gamma}_t = \Gamma_0 = \Gamma (A_t)$. The enveloped or coarse-grained
phase volume $\bar{\Gamma}_t$ grows approximately as
(\ref{eq:eq3.16}).
The number of states in $\bar{\Gamma}_t$ depends on the definition
of a state in the enveloped phase volume. Let one state occupies an
elementary volume $\Delta\Gamma$. Then the number of states that occupy
the volume $\bar{\Gamma}_t$ is simply
\begin{equation}
\label{eq:eq3.18}
{\cal N} (t; \Delta\Gamma ) = \bar{\Gamma}_t /\Delta\Gamma =
(\Gamma_0 /\Delta\Gamma ) \exp (ht)
\leq \Gamma (M) /\Delta\Gamma .
\end{equation}

Let us emphasize that this interpretation stops working when
$t \gg 1$ (in fact, when
$t> \frac{1}{h}\log \left(\frac{\Gamma (M)}{\Gamma_0}\right)$).


Now instead of the $\epsilon$-separated trajectories we can introduce
$\Delta\Gamma$-separated ones,
associated to a partitioning of $\Gamma (M)$ into
elementary cells. Two trajectories will be
$\Delta\Gamma$-separated over the time interval $t$ if they do not stay
at the same cell $\Delta\Gamma_j$ during $t$.
If the volume of an elementary cell is
$\Delta\Gamma = \epsilon^b$, then
\begin{equation}
\label{eq:eq3.19}
{\cal N} (t,\Delta\Gamma ) \equiv
{\cal N} (t,\epsilon ) = (\Gamma_0 /\epsilon^b ) \exp (ht)
\end{equation}
and we arrive at the connection
\begin{equation}
\label{eq:eq3.20}
C_{\epsilon ,t} (A_0 ) = {\rm const} \ {\cal N}(t,\epsilon )
\end{equation}
We also can introduce an entropy $S(t,\Delta\Gamma )$ for the
$\Delta\Gamma$-separated states, i.e.
\begin{equation}
\label{eq:eq3.21}
S(t,\Delta\Gamma ) \equiv \ln N(t,\Delta\Gamma ) =
\ln C_{\epsilon ,t} (A_0 )
+ {\rm const}
\end{equation}
with a condition for the states-separation
\begin{equation}
\label{eq:eq3.22}
\Delta\Gamma = \epsilon^b \ .
\end{equation}
It is essential that the entropy $S(t,\Delta\Gamma )$ is defined
relatively to a definition of the elementary phase volume
$\Delta\Gamma$ which depends on the system and the type of the evolution
process. Non universality of the entropy $S(t,\Delta\Gamma )$ will be
more evident in the next sections
where the algebraic complexity will be
considered. Here we only would like to mention that partitioning of
phase space depends on the level of information we want to ignore
in the description of dynamics and on the level of information we would
like to keep about system trajectories.

A few other comments also are useful. $\Delta\Gamma$-partitioning
resembles a procedure of coarse-graining, which is typical in
statistical mechanics. $\Delta\Gamma$-partitioning is not the same as
$\epsilon$-partitioning since trajectories from neighboring cells can be
arbitrarily close to each other during an arbitrarily large $t$, although
we will never know it. At the same time there are common features between
$\Delta\Gamma$- and $\epsilon$-partitioning. The formal expressions
(\ref{eq:eq3.13}) for $C_{\epsilon ,t} (A)$ and (\ref{eq:eq3.19}) for
$N(t,\Delta\Gamma )$ are the same up to a const (compare to
(\ref{eq:eq3.20}),(\ref{eq:eq3.22})). Both definitions have the same
limitation for the bounded Hamiltonian dynamics:
\[
C_{\epsilon ,t} (A) \leq \Gamma (M) /\epsilon^b
\]
\begin{equation}
\label{eq:eq3.23}
{\cal N} (t;\Delta\Gamma )\leq \Gamma (M) /\Delta\Gamma .
\end{equation}
The constraints (\ref{eq:eq3.23}) do not depend on time. Thus, it follows
from (\ref{eq:eq3.23}) the existence of
\begin{equation}
\label{eq:eq3.24}
t_{\max} = t_c (\epsilon ) \ {\rm or } \
t_{\max} = t_N (\Delta\Gamma )
\end{equation}
such that trajectories or their segments not separated during the
$t_{\max}$ will be non-distinguishable. This property of the definitions
(\ref{eq:eq3.13}),(\ref{eq:eq3.19}) eliminates a significant part of the
dynamics with non-uniform mixing in phase space.

\section{Complexity functions}
\label{sec:sec4}

The goal of this chapter is to introduce a new definition of a complexity
function (CF) rather than just complexity, in order to be able to
characterize a system with at least two different time scales of a
``complex'' dynamics.

\subsection{Definitions}
\label{sec:sec4.1}

As before, we deal with area-preserving dynamics of systems in a metric
phase space $M$ endowed with a distance $dist$ and discrete or continuous
time $t$. The dynamical system $f^t : \ M\rightarrow M$ defines a distance
$dist(f^t x,f^t y)$ between two trajectories at time $t$ that were initially
at points $x,y \in M$. One can also introduce a natural length
$\ell^t = \ell (x,f^t x) = \ell (x,x^t )$ along the trajectory initially
at $x$. We shall need the following definition:

Two trajectories with initial points $x,y \in A$
will be $(\epsilon ,t)$-indistinguishable if
\begin{equation}
\label{eq:eq4.1}
d_{\tau} (x,y) = dist(f^{\tau} x,f^{\tau} y) < \epsilon , \ \  \ \
0 \leq \tau< t.
\end{equation}

We now introduce a notion of complexity that is based on the verification
of divergence of trajectories from fixed several ones. We start with a
definition of local complexity.

Consider a small domain $A \subset M$ with diameters $\delta_A$, $\delta_M$
where $\delta_A\ll\delta_M$, and fix some number $\epsilon : \ \delta_A \ll\epsilon\ll\delta_M$.

Let us pick a point $x_0\in A$ and call the corresponding trajectory the
basic one.
A set $Q_N=\{x_k\in A\}_{k=1}^N$ is said to be locally 
($\epsilon$, t)-separated if

\begin{itemize}
\item[(i)] For every $x_k$ there is $0\leq \tau_k \leq t$ such that
\begin{equation}
dist (f^{\tau_k} x_k, f^{\tau_k} x_0)\geq \epsilon
\end{equation}
and
\begin{equation}
dist (f^{\tau} x_k, f^{\tau} x_0)< \epsilon, \quad 0\leq \tau < \tau_k,
\end{equation}
\item[(ii)] for every pair ($k$, $k'$), $1\leq k$, $k'\leq N$, one has
\begin{equation}
\label{distance}
dist (f^{\tau_k} x_k, f^{\tau_{k'}} x_{k'})\geq \epsilon.
\end{equation}
\end{itemize}

If (\ref{distance}) is not valid then a pair of trajectories
corresponding to the pair $(k,k')$ is $\epsilon$-indistinguishable,
and it should be treated as one trajectory during the time $t$.

{\bf Definition 2.} The number
\begin{equation}
C(\epsilon,t,x_0,A):=\max \{ N \; | \;
Q_N \textup{ is locally (}\epsilon\textup{, t)-separated}\}
\end{equation}
is called the local complexity. As a function of $t$ it is said to be
the local complexity function.

A set $Q_N$ is called ($\epsilon$, t)-optimal if it is locally 
($\epsilon$, t)-separated and $N=C(\epsilon,t,x_0,A)$.

\vspace{3ex}

It is simple to see that  $C(\epsilon,t,x_0,A)\leq \Gamma(M)/\epsilon^b$ and
\begin{equation}
C(\epsilon,t',x_0,A)\geq C(\epsilon,t,x_0,A)\quad \textup{if} \quad t'\geq t.
\end{equation}

\begin{figure}[ht]
\includegraphics[width=0.5\textwidth]{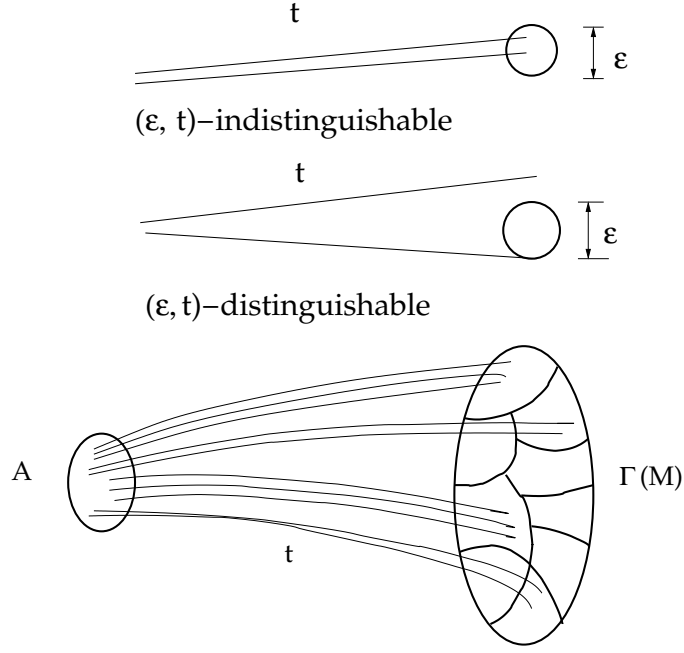}
\caption{Example of the evolution of 10 trajectories from the set $A$ of the size << $\epsilon$ during time $t$. The optimal (maximal) number of $(\epsilon, t)$--separated trajectories $N=9$. Diameter of all 9 domains of $\Gamma(M)$ is $\epsilon$. Te initial bunch $Q_N$ has 10 trajectories with $N_0=C(\epsilon, t; x_0, A)=4$ of the $(\epsilon, t)$--separated ones. Two top examples show the process of evolution that makes a pair of trajectories (in)distinguishable.}
\label{new1}
\end{figure}

So, if we are interested in the separation of a bunch $Q_N$ of $N$ trajectories
with initial points $x_j\in Q_N$, $j=1,..., N,$ and their evolution 
$x_j(t)=f^t(x_j)$, then after time $t$ we find out that there are 
$N_0=N_0(\epsilon,t,Q_N)$ $\epsilon$-separated (from the basic one)
trajectories- denote this set by $Q_{N_0}$- and $N-N_0$
indistinguishable (from the basic one) trajectories (see Fig.~\ref{new1}).

If $Q_{N_0}$ is locally separated then we obtain an estimate
\begin{equation} \label{N0x0}
N_0(\epsilon,t,Q_N)\leq C(\epsilon,t,x_0,A).
\end{equation}
If, in addition, $Q_{N_0}$ is ($\epsilon$, t)-optimal then
\begin{equation} \label{N0x01}
N_0(\epsilon,t,Q_N)= C(\epsilon,t,x_0,A).
\end{equation}
From physical point of view it is natural to put the following
restriction
\begin{equation}\label{volume}
N_0(\epsilon,t,Q_N)\leq {\cal N} =\frac{\Gamma(M)}{\epsilon^b}.
\end{equation}
See Subsection D of Section 3.

Now consider a partitioning of the full phase volume $M$ by a set of
domains $A_k : \ M= \cup_k A_k$, and in each $A_k$ select points
$x_0^{(k)}$ that make a set of basic trajectories
$B_k : \ x_0^{(k)} \in B$. Let us assume for the sake of definiteness
that 
\begin{equation}\label{kk'}
dist(x_0^{(k)},x_0^{(k')})> 2\epsilon, \quad k\neq k'.
\end{equation}
Because of (\ref{kk'}), every two points belonging to
$\epsilon$-neighborhoods of different basic points are
($\epsilon$, t)-separated for any $t\geq 0$.

We consider the finite set $Q_N= \{x_j\}_{j=1}^{N}$, $N\gg 1$.
Every point in $Q_N$ belongs to the one of sets $A_k$, so we have
a partition $Q_N=\cup_k Q_N^{(k)}$, $ Q_N\cap A_k=Q_N^{(k)}$.

We call the set $Q_N$ semi-locally ($\epsilon$, t)-separated, or
simply   ($\epsilon$, t)-separated, if
every $Q_N^{(k)}$ is locally ($\epsilon$, t)-separated
(with respect to the basic point $x_0^{(k)}$).

{\bf Definition 3.} The number
\begin{equation}
C(\epsilon,t,B,\{A_k\}):=\max \{ N \; | \;
Q_N \textup{ is  (}\epsilon\textup{, t)-separated}\}
\end{equation}
is called the semi-local complexity, or simply complexity. As a
function of $t$ it is said to be 
the complexity function.

\vspace{3ex}

If we choose $N^{(k)}$ initial points at each set $A_k$ and
consider $N_0^{(k)}=N_0^{(k)}(\epsilon,t,x_0^{(k)},A_k)$ of them
corresponding to trajectories ($\epsilon$, t)-separated from the
basic one, then we may form the sum
\begin{equation}\label{N0B}
\sum_{A_k}  N_{0}^{(k)} (\epsilon ,t, x_0^{(k)},A_k )\leq
C(\epsilon,t,B,\{A_k\}).
\end{equation}
If these $N_0^{(k)}$ points form an optimal set for every $k$ then
this inequality becomes the equality:
\begin{equation}\label{N0B1}
C(\epsilon,t,B,\{A_k\})=
\sum_{A_k}  N_{0}^{(k)} (\epsilon ,t, x_0^{(k)},A_k ).
\end{equation}

The complexity functions $C(\epsilon,t,x_0,A)$ and
$C(\epsilon,t,B,\{A_k\})$ show a level of time-proliferation of
$\epsilon$-separated trajectories from the initial set of a large number
of indistinguishable points.

If we are interested in the only typical (for some measure) orbits we may
adjust definitions above to this situation by assuming that $Q_N$ always
consists of typical points for this measure and the points $x_0^{(k)}$
are also typical (see below).

One of the main points we are interested in is behavior of complexity
functions in a neighborhood of the sticky set. It is well-known that
a ``standard'' trajectory in chaotic sea behaves in the intermittent way:
after relatively short chaotic burst it is attracted to the sticky set
for a long time, then comes back to mixing part of the chaotic sea, etc.
If our consideration is restricted to a neighborhood of one (or several)
basic orbit, then fast separated pieces of orbits correspond to a mixing type
of behavior and their initial points are situated much ``far'' from
the sticky set than initial points of slow-separated pieces of orbits.
By using this observation, we may eliminate fast-separated points
(see the next section) i.e., in fact we may choose such initial points in 
$Q_N$ which practically belong to the sticky set. In other words, we may
in principle calculate local and semi-local complexity of a measure
concentrated on the sticky set.

More rigorously, assume that an invariant measure $\mu$ is given. Then in
definition 2 we consider only sets $Q_N=\{x_j\}_{j=1}^N$ such that
\begin{itemize}
\item[(i)] $Q_N$ is locally ($\epsilon$,t)-separated;

\item[(ii)] $x_j \in supp(\mu)\cap A$, $j=1,...,N,$
\end{itemize}
(where $ supp(\mu)$ is the smallest closed set of full measure i.e.,
the set on which the measure $\mu$ is concentrated).
The maximal number of elements in such sets $Q_N$ will be called the
local ($\epsilon$,t)-complexity of measure $\mu$. We denote it by
 $C_{\mu}(\epsilon,t,x_0,A)$.

Similarly if in definition 3 we consider only sets $Q_N$ containing
points from $ supp(\mu)$, then we obtain the semi-local
 ($\epsilon$,t)-complexity of measure $\mu$,
$C_{\mu}(\epsilon,t,B,\{A_k\})$

It is useful to introduce the following quantity
\begin{equation}
\label{eq:eq4.8}
P_\mu(\epsilon ,t;\Delta t,x_0) = {1\over {N} }
[C_\mu(\epsilon ,t+\Delta t,x_0,A) -
C_\mu(\epsilon ,t,x_0,A)] \approx p_\mu (\epsilon ,t,x_0) \Delta t,
\end{equation}
where $N=C_\mu(\epsilon ,t,x_0,A)$. This quantity
gives a probability to diverge by distance $\epsilon$ from the basic orbits
during the time
interval $[\tau ,\tau + \Delta t]$, and $p_\mu(\epsilon ,t,x_0)$ is the
corresponding probability density function.

Similarly,
\begin{equation}
P_\mu(\epsilon ,t;\Delta t,B) = {1\over {N} }
[C_\mu(\epsilon ,t+\Delta t,B,\{A_k\}) -
C_\mu(\epsilon ,t,B,\{A_k\})] \approx p_\mu (\epsilon ,t,B) \Delta t
\end{equation}
gives the probability to diverge from basic orbits going through $B$
during the time interval  $[\tau ,\tau + \Delta t]$,
and $p_\mu(\epsilon ,t,B)$ is the corresponding probability density
function.

\subsection{Calculation of the local complexity function}
\label{sec:sec4.2}


From now, we will choose initial points $x_j$ in $Q_N$ in such a way that
the distances between $x_j$ and the basic point $x_0$ are the same for all $j$,
and we denote it by $\delta$ Moreover, for the sake of simplicity we
omit the argument $A$ in $C(\epsilon,t,x_0,A)$ and $\{A_k\}$ in 
$C(\epsilon,t,B,\{A_k\})$. So 
$C(\epsilon,t,x_0)=C(\epsilon,t,x_0,A)$ and 
$C(\epsilon,t,B)=C(\epsilon,t,B,\{A_k\})$.

As usual in numerical simulations we will assume that randomly chosen
points $x_0$, $x_0^{(k)}$ and $Q_N$ are typical with respect to
some measure $\mu$ we are interested in.

The values of $N_0$ in (\ref{N0x0}), (\ref{N0x01}) and in
 (\ref{N0B}), (\ref{N0B1}) depend on the choice of $\delta$,$\epsilon$, $x_0$
(or $B$) and $Q_N$.



The smaller is $\delta$, the longer $t$
should be considered until the maximal values of $C(\epsilon ,t,B)$ or
$C(\epsilon ;t,x_0)$ will be achieved. This makes the limit
$\delta\rightarrow 0$ fairly simple. Understanding a way to work with the
parameter $\epsilon$ is more complicated. Consider one trajectory $x^t$
that starts at $x$ and has a natural length $\ell = \ell (x,x^t )$. Let $t$
be fairly big and select a set of points $x_k$ along a trajectory and,
approximately, almost uniformly distributed. We can operate with points
$x_k$ in the same way as with the basic points $x_0^{(k)}$ in
(\ref{N0B1}). As a result, we obtain the quantity
\begin{equation}
\label{eq:eq4.11}
\bar{C} (\epsilon ;t,x_k ) = N_0 (\epsilon ;t,x_k )
\leq {\Gamma (A) \over \epsilon^b}
\end{equation}
where points $x_k$ belong to the same trajectory. That means that while
$C(\epsilon ;t,x)$ characterizes the $\epsilon$-divergence from $x$ during
$t$, $\bar{C} (\epsilon ;t,x_k )$ characterizes the $\epsilon$-divergence
from $x_k$: points $x$ are taken in different places of the phase space and
points $x_k$ are taken along the only trajectory. It is natural to believe
that
for an appropriately typical set of $x\in M$ and of
$x_k \in \ell (x,x^t )$ and fairly large $t$, the equality
\begin{equation}
\label{eq:eq4.12}
C(\epsilon ;t,x) = \bar{C} (\epsilon ;t,x) , \ \  \
t\rightarrow\infty
\end{equation}

holds where subscripts $k$ are omitted. This equality  can be treated
as an analog of the ergodic theorem.

\begin{figure}[h]
\includegraphics[width=0.5\textwidth]{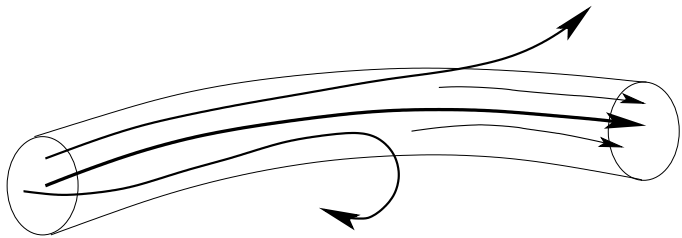}
\caption{$\epsilon$-separation from the basic trajectory.}
\label{new2}
\end{figure}

A corresponding simulation for $\bar{C} (\epsilon ;t,x)$ was performed in
[LZ] for a system of tracer dynamics in the field of point vortices. The
basic trajectory was created by a tracer, and a few
``host tracers'' were considered within a small distance $\sim\delta$ from
the basic trajectory. Each time, when a host tracer moves at a distance
$\epsilon$ from the basic one, it was removed and replaced by a new host
tracer at a distance $\delta$ from the basic one (Fig.~\ref{new2}).

The scheme of calculation of $\bar{C} (\epsilon ;t,x)$ is similar to one used
for calculation of the Lyapunov exponents, except for some details:

\begin{enumerate}
\item[(a)]
the value of $\epsilon$ was much less than the distance typically used to
evaluate the Lyapunov exponents;
\item[(b)] trajectories of some tracers were $\epsilon$-separated after
a very long time. Just these trajectories correspond to events of our main
interest and their statistics was collected.
\item[(c)] the scheme of obtaining of $\bar{C} (\epsilon ;t,x)$ provides
simultaneously two different distributions: ``probability''
$\bar{C} (\epsilon ;t,x)$ to have $\epsilon$-separation at time $t$,
and $\bar{C} (\epsilon ;\ell^t ,x)$ as a ``probability'' to have
$\epsilon$-separation after the travel over natural length $\ell^t$ along
the trajectory. Both of these probabilities cannot be obtained by a simple
transformation of variables $t=t(\ell )$ since the trajectories can be of
a fractal type
and the variables may not be transformable. These necessitates to use more
general distribution than $\bar{C} (\epsilon ;\ell^t ,x)$ or
$\bar{C} (\epsilon ,t,x)$. This will be fulfilled in the following section.
\end{enumerate}

\subsection{Complexity function and exit time distribution}
\label{sec:sec4.3}

The heuristic consideration of the previous section can be formalized in a
more accurate way and linked to the distribution function of exit time. Let
us return back to the set $A$ with a basic trajectory that starts at
$x \in A$, and let $A_{\epsilon ,t}$ be the set of $\epsilon$-separated
during the time interval $(0,t)$ trajectories.
Assume that the system $(f^t ,M)$ is transitive and the semi-orbit
\begin{equation}
\label{eq:eq4.13}
{\cal L} = \cup_{\tau = 0}^{\infty} f^{\tau} x
\end{equation}
is dense in $M: \ \bar{\cal L} = M$. It follows from that:

\noindent
{\bf Proposition 1.} For any $t\gg 1$ there is $\delta > 0$ such that for
any $y\in M$ there exists $x\in {\cal L}$ such that if dist$(x,y) < \delta$,
then dist$(x^{\tau} ,y^{\tau} ) \leq \epsilon$, $0\leq \tau \leq t$.

Proposition 1 states the existence of the indistinguishable trajectories,
and the $\epsilon$-separation of orbits can be interpreted as an
exit time event of $y^t$ from the $\delta$-vicinity of ${\cal L}$,
$(\delta\ll\epsilon )$. Let ${\cal L}_t$ be a piece of the
basic trajectory $(x,x^t )$ with the length $\ell_t = \ell (x,x^t )$.

Given this basic orbit ${\cal L}_t$, one may study a distribution of pairs
with respect not only to exit times but also to positions in space
measured by a coordinate $s$
along the orbit $\cal L$. Indeed, let us measure
the position of the point $f^{\tau} x$ on $\cal L$ by the coordinate
$s$ equal to the length   $\ell_{\tau} = \ell (x,x^{\tau} )$ of the piece
of orbit ${\cal L}_{\tau}$ with the initial
point $x$ and the final point $f^{\tau}_x$. Because of the Proposition 1,
for an arbitrary point $y \in M$ the ball $B_{\delta} (y)$ of radius
$\delta$ centered at $y$ contains at least one point
$x \in{\cal L} $.
One may characterize the trajectory $(y,y^t )$ by the same parameter $s$
as $(x,x^t)$.

Let $A_{\epsilon ,t}$ be a $(\epsilon, t)$-separated set and
$x_j \in A_{\epsilon ,t}$. Fix a coordinate $s_j$ of $x_j$ by choosing the
initial point at $x_j$ to start counting $s_j$.
Now, introduce into consideration the pair $(\tau_j, s_j )$ where
$\tau_j = \tau (x_j )$ is the first time for which
$dist(f^{\tau_j}x_j ,f^{\tau_j} x ) = \epsilon$, i.e. the exit time
from the $\epsilon$-neighborhood of the basic trajectory
${\cal L}_t$, and
$s_j := s(f^{\tau_j} x_j )-s(x_j )$, the length of the piece of the
orbit between the initial and final points.

For two different $x_j$ and $x_k$ that are close to the basic $x$ and escape
from the $\epsilon$-vicinity of $x^t$ at close times $\tau_j$ and
$\tau_j^{\prime}$ one can expect that the orbit segment's length
$s_j ,s_j^{\prime}$ will also be close. But if we take another $\bar{x}$
as a basic one, other two initial points $\bar{x}_j ,\bar{x}_j^{\prime}$ in
the $\delta$-vicinity of $\bar{x}$ such that the $\epsilon$-escape time
for them will be correspondingly $\tau_j$, $\tau_j^{\prime}$, i.e. the
same as for the initial basic point $x$, then it could not be that
$\bar{s}_j ,\bar{s}_j$ are the same. This property is a result of the
non-uniformity of behavior of orbits in the phase space and, speaking further,
it is a result of the fractal space-time properties of trajectories.

\subsection{Flights complexity function (FCF)}
\label{sec:sec4.4}

Consider a small domain $A\subset M$, a typical point $x\in A$ and a typical
close point $y\in A$ at $\delta$-distance from $x$. The
$\epsilon$-separation of the pair $(x,y)$, $\epsilon \gg\delta$ occurs at
some time $t$ and distance $s = \ell^t = \ell (x,x^t )$.
Here we omitted $\epsilon$ as an argument of $l$. Furthermore, let us
remark that the distance $s$ depends on $y$. But we neglect this
dependence assuming implicitely that for a typical $y$ at the
$\delta$--distance from $x$ the value of $s$ is asymptotically the same.

The distance $s$ of the $(\epsilon ,t)$-separation is said
to be the length of a flight, i.e. a length of the path that two nearby
trajectories are flying together.

\vspace{3ex}

Let us treat $x$ as the basic point and consider the set
$Q_N=\{x_k\}_{k=1}^N$ of points $\delta$-close to $x$. This set is called
locally ($\epsilon$, t, s)-separated if it is 
locally ($\epsilon$, t)-separated (see Subsection A)
and moreover $s_k<s$ where $s_k=\ell (x,x^{\tau_k} )$.

\noindent
{\bf Definition 4.} The number
\begin{equation}
C(\epsilon, t, s, x)= \max \{N \; | \; Q_N
\textup{ is locally ($\epsilon$, t, s)-separated}\}
\end{equation}
is called the local  ($\epsilon$, t, s)-complexity.
As a function of $t$, $s$, it is called the local flight complexity
function (FCF).

\vspace{3ex}

Similarly to the definition of $C(\epsilon ,t, B)$ in (\ref{N0B}),
(\ref{N0B1}) we can consider a
collection of flights and their lengths and time intervals from different
domains $A_k : \ M = \cup_k A_k$. As a result we have the semi-local
FCF
\begin{equation}
\label{eq:eq4.14}
C(\epsilon ,t,s,B) = {\sum_{A_k}} C(\epsilon ,t,s,x_0^{(k)})
\end{equation}
where $x_0^{(k)}\in A_k$ are basic points.

As in Subsection A we may choose $N^{(k)}$ initial points in $A_k$ for
any $k$, select the maximal ($\epsilon$, t, s)-separated subset
consisting of $N_0^{(k)}(\epsilon ,t,s)$ points and form the sum
$$
{\sum_{A_k}} N_0^{(k)} (\epsilon ;t,s)\leq C(\epsilon ,t,s,B).
$$
The sum is equal to the semi-local FCF if initial points form an optimal set.

As in Subsection C, let us assume that there exists a trajectory ${\cal L}$
dense in our phase space $M$. Then, as it was said there, every piece of a
trajectory $(y,y^t)$ lies in an $\epsilon$-vicinity of a piece
$(x,x^t)$, $x\in {\cal L}$, and, thus has the coordinate $s=\ell (x,x^t)$.
It allows us to extend the notion of ($\epsilon$, t, s)-complexity as
follows.

We say that a set $Q_N=\{x_k\}_{k=1}^N$ is ($\epsilon$, t, s)-separated
if it is ($\epsilon$, t)-separated and $s_k\leq s$ where $s_k$ is the
$s$-coordinate of the piece $(x_k,x_k^t)$.

\noindent
{\bf Definition 5.} The number 
\begin{equation}
C(\epsilon, t, s )= \max \{N \; | \; Q_N
\textup{ is ($\epsilon$, t, s)-separated}\}
\end{equation}
is called the flight complexity.
As a function of $t$, $s$, it is called the flight complexity
function.

\vspace{3ex}



This quantity  has
a simple meaning: let us take a large enough number ${\cal N}_{\delta}$ of
$\delta$-close pairs within $M$ with a fairly typical distribution of the
initial conditions $x_k \in M$, $k=1,\ldots ,{\cal N}_{\delta}$.
$C(\epsilon ;t,s)$
is the number of
trajectories that are mutually $\epsilon$-separated $(\epsilon
\gg\delta )$ during time $t$
at distance $s$
Therefore  $C(\epsilon,t+\Delta t, s +\Delta s)-C(\epsilon,t, s)$
is the number of
trajectories that are $\epsilon$-separated during time
$t\in (t,t+\Delta t)$
at distance $s\in (s,s+\Delta s)$.

One can introduce the function
\begin{equation}
\label{eq:eq4.15}
p(\epsilon ;t,s)  \cong {1\over {N}} \ {1\over\Delta t} \
{1\over \Delta s} [C(\epsilon ;t+\Delta t, s +\Delta s) -
C(\epsilon , t,s)]
\end{equation}
where $N=C(\epsilon , t,s)$, is called
the density of the FCF.

\vspace{3ex}

The main goal of this paper is to connect the notion of complexity with
the space-time local instability properties of systems. A necessity of such
a notion appears due to the specific features of Hamiltonian systems to have
chaotic dynamics strongly non-uniform in phase space and strongly
intermittent in time, and particularly due to the existence of dynamical
traps [Za2].

Let us consider again a small domain $A$ and pick some $N$ trajectories
in $A$ with a characteristic distance $\delta$ between them. Due to
the instability these $N$ trajectories will fill at time $t$ an enveloped
phase volume $\Gamma_{\rm env} (t;A,\delta )> \Gamma (0,A,\delta )$
because of the larger distance between trajectories. After coarse-graining
of $\Gamma_{\rm env} (t;A,\delta )$ we arrive at
$\bar{\Gamma} (t,A) \sim \Gamma_{\rm env} (t; A,\delta )$ but without
empty space (``bubbles'') presented in
$\Gamma_{\rm env} (t; A,\delta )$.
This is a typical physical situation of growth of the coarse-grained phase
volume and the problem is reduced to the estimate of the growth. The presence
of dynamical traps makes the growth of $\bar{\Gamma} (t,A)$ sensitively
dependent on all its parameters $t$, $A$ and $\delta$ or, in other words,
expanding of $\bar{\Gamma}$ depends on the initial coordinate and
observational time. For example, if $A$ is taken in the stochastic sea,
$\bar{\Gamma} \sim \exp (\hbar t)$ but if $A$ is taken inside a dynamical
trap, $\bar{\Gamma} \sim t^{\mu}$.

We need to make the only step to arrive to our definition of complexity.
Indeed, one can consider the phase volume partitioning by  elementary
cells $\Delta\Gamma$ and introduce the complexity
\begin{equation}
C(\epsilon ,t;x) = {\rm const.} \ \bar{\Gamma} (\epsilon ,t;x)
\label{A.1}
\end{equation}
where we use $x$ as a coordinate of $A$ and $\epsilon$ defines a diameter
of the coarse-graining and does not enter into any physical results.

The density of the FCF (49) can now be considered as a distribution function
of having displacement $s$ at a time instant $t$ and the corresponding
entropy can be easily introduced (see the next section).

One more useful connection appears from the expression
\begin{equation}
P_{\rm tr} (\epsilon ,t;A) = \frac{1}{\Gamma (A)} \int_0^t
dt \ p(\epsilon ,t;A) \label{A2}
\end{equation}
that is a probability to be trapped in a tube of the diameter $\epsilon$
supported by the domain $A$ during time $t$.

Respectively
\begin{equation}
P_{\rm esc} (\epsilon ,t;A) = 1-P_{\rm tr} (\epsilon ,t;A)
\label{A.3}
\end{equation}
is a probability to exit from the tube during some time $> t$, and for
$t\rightarrow\infty$ and $P_{\rm esc}\sim 1/t^{1+\beta}$ we arrive at
\begin{equation}
P_{\rm rec} (t) \sim 1/t^{\gamma}
\label{A.4}
\end{equation}
of the probability density to return to any small domain $\cal A$ with
\begin{equation}
\gamma = \min\beta (A) +1
\label{A.5}
\end{equation}
and no dependence on $\cal A$. Due to the Kac lemma $\gamma > 2$ for the
bounded Hamiltonian dynamics.

In this chapter we considered the case of continuous time. But
all ideas, definitions and results  can be generalized
to the case of discrete time. One can choose different analogues of
the length of the piece of an orbit- for example in an euclidean space
one may consider the length of union of segments joining consecutive
points of the orbit. The main ideas are independent of the choice of
the definition of the lengths.

\section{Entropy}
\label{sec:sec5}

The function $C(\epsilon ;t,s)$ characterizes local $\epsilon$-divergence of
trajectories and it can be considered as a new characteristic of the
dynamics. 
Its main role is to describe evolution of a
typical pair of orbits taken apart of a small distance $\delta$. Let us
show how the $C(\epsilon, t,s)$-complexity works.

\subsection{Complexity and entropy}
\label{sec:sec5.1}

Following a general physical approach let us define
\begin{equation}
\label{eq:eq5.1}
S(\epsilon ,t,s) = \ln C(\epsilon ,t,s)
\end{equation}
as $(\epsilon , t,s)$-entropy of the dynamics since $C(\epsilon ;t,s)$ is a
number of $(\epsilon ;t,s)$-separated states. Physically speaking, due to
(\ref{eq:eq4.11}) the
number of separated states cannot be more than
${\cal N} = \Gamma (M)/\epsilon^b$ (compare to (\ref{volume}), i.e. for the
ergodic dynamics
\begin{equation}
\label{eq:eq5.2}
S(\epsilon ,t,s) \leq \ln {\cal N} .
\end{equation}
Due to  a mixing property in phase spaces or, more specifically, due to an
instability and separation of any initially close orbits, one can expect the
growth of $S(\epsilon ;t,s)$ with time until the max $S=\ln {\cal N}$ is
reached. The definition (\ref{eq:eq5.1}) permits to estimate from
$C(\epsilon ,t,s)$ some fine properties of the evolution of
$S(\epsilon ,t,s)$ as a function of time $t$ and the length of separation
$S$.

\subsection{Examples}
\label{sec:sec5.2}

For the case of the Anosov-type systems
\begin{equation}
\label{eq:eq5.3}
s = s(t) = s(0) e^{ht} ,
\end{equation}
i.e. $s$ has a ``good'' dependence on time and it may be expunged from
$C(\epsilon ,t,s)$. Then
\begin{equation}
\label{eq:eq5.4}
C(\epsilon ;t,s) \rightarrow C(\epsilon ,t) = \left( {\delta\over\epsilon}
\right)^b e^{ht}
\end{equation}
where $s(0) = \delta$ and (\ref{eq:eq5.4}) coincides with
(\ref{eq:eq3.13}) (and the arrow means ``is completely defined by'').
Similarly
\begin{equation}
\label{eq:eq5.5}
S(\epsilon ;t,s) \rightarrow S(\epsilon ,t) = \ln C(\epsilon ,t) = b\ln
{1\over\epsilon} + ht  + {\rm const}
\end{equation}
and we arrive to the standard expressions
\[
h = {\partial S(\epsilon ,t) \over \partial t} = {\rm const}
\]
\begin{equation}
\label{eq:eq5.6}
b = {\partial S(\epsilon ,t) \over \partial\ln (1/\epsilon )}= {\rm const}
\end{equation}
which deliberately are written in the form of partial derivatives.

Consider fully opposite case of an integrable system in
the annulus
$K = \{ (\theta ,r)$, $0 \leq \theta\leq 2 \pi$, $1 \leq r \leq 2\}$,
defined by the equation
\begin{equation}
\bar{r} = r \ , \ \ \ \
\bar{\theta} = \theta +r, \ \ \ \ {\rm mod} \ 2\pi
\label{eq:eq5.7}
\end{equation}
i.e. the map $T: \ K \rightarrow K$ is
$T(r,\theta ) = (r,\theta +r)$.
Let us calculate $C_{\epsilon ,t} (A)$
where $A$ is the interval $\theta = \theta_0$, $1 \leq r \leq 2$. Since
$T^t (r,\theta_0 ) = (r,\theta_0 + rt \ {\rm mod} \ 2\pi )$, the
segments $\ell_t (r_0 ,\vartheta_0 )$, $\ell_t (r_1 ,\theta_0 )$ are
$(\epsilon ,t)$-separated iff $|r_0 t-r_1 t| \geq \epsilon$, i.e.
$|r_0 - r_1 | \geq \epsilon /t$. The maximal number of such points on the
interval [1,2] is $1/\epsilon \cdot t$, and hence
\begin{equation}
\label{eq:eq5.8}
C_{\epsilon ,t} (A) = {t\over\epsilon }
\end{equation}

Since the number of $\epsilon$-different sections $\theta = \theta_0$
on the interval $[0,2\pi ]$ is ${2\pi\over\epsilon}$, then one has
\begin{equation}
\label{eq:eq5.9}
C_{\epsilon ,t} (K) = 2\pi\cdot\epsilon^{-2}\cdot t
\end{equation}
Thus, integrable nonlinear system on the annulus is a system with linear
growth of complexity. By the way, $b=2$ is the dimension of the annulus.

The corresponding entropy is
\begin{equation}
\label{eq:eq5.10}
S(\epsilon ,t) = \ln t - 2\ln\epsilon + {\rm const}
\end{equation}
and
\begin{equation}
\label{eq:eq5.11}
{\partial S (\epsilon ,t) \over \partial t} \rightarrow 0 , \ \  \
(t\rightarrow\infty )
\end{equation}
This shows that in a polynomial case it is possible to consider the expression
\begin{equation}
\label{eq:eq5.12}
{\partial S (\epsilon ,t) \over \partial \ln t} 
\end{equation}
(equals to 1 for (\ref{eq:eq5.10})) which provides the information on the
frequency (equals 1)
of the rotation in the annulus.

There are many other less trivial systems for which growth of complexity
is subexponential (billiards in polygons, interval exchange transformations,
etc.). 


\subsection{Polynomial complexity and anomalous transport}
\label{sec:sec5.3}

Complexity defined in (\ref{eq:eq5.4}) grows exponentially with time
reflecting an existence of a positive Lyapunov exponent and exponential
divergence of trajectories in phase space. This is not the case for zero
Lyapunov exponent systems and for diffusional type processes. To consider
large scale processes such as the diffusion
for systems of that kind,
let us use the partition
function ${\cal N}(t,\Delta\Gamma )$
\begin{equation}
\label{eq:eq5.13}
{\cal N}(t,\Delta\Gamma ) = {\rm const.} \ C(\epsilon ,t)
\end{equation}
with a normalization constant that can be chosen in such a way that
\begin{equation}
\label{eq:eq5.14}
{\cal N} (t,\Delta\Gamma )\rightarrow 1 , \ \ \ \ t\rightarrow\infty
\end{equation}
In numerous cases the process of mixing in phase space and the corresponding
separation of trajectories has algebraic dependence on time. Particularly,
it is related to the dynamics with the fractal time [$\ldots$] and the
diffusional type of processes. For this case, the complexity is polynomial in time
and
\begin{equation}
\label{eq:eq5.15}
{\cal N} (t,\Delta\Gamma ) = {\rm const.} \cdot (t/t_0 )^{\beta} =
{\rm const.} \cdot C(\epsilon ,t)
\end{equation}
with some exponent $\beta$ and characteristic time scale $t_0 \ll t$.
As in (\ref{eq:eq5.12}), the entropy growth rate can be defined in the
$\ln t$ scale
\begin{equation}
\label{eq:eq5.16}
{\partial S(\epsilon ,t) \over \partial\ln t} =
{\partial\ln C(\epsilon ,t) \over \partial\ln t} = \beta
\end{equation}
and the exponent $\beta$ characterizes the corresponding rate of the entropy
growth. As it appeared in [$\ldots$], chaotic dynamics of real Hamiltonian
systems does not display, in general, exponential growth of the complexity,
and the number $\cal N$ of $\epsilon$-separated orbits grows asymptotically
as polynomial in time and the length of separation $s = \ln (x,x^t )$. In
this case the complexity can be written as
\begin{equation}
\label{eq:eq5.17}
C(\epsilon ;t,s) = (s_0 /s)^{\alpha} (t/t_0 )^{\beta}
g(\epsilon ;s,t)
\end{equation}
where $s_0 ,t_0$ are characteristic scales of length and time,
$\alpha ,\beta > 0$, and $g(\epsilon ;s,t)$ is a slow varying function.
We will be interested in the limit
\begin{equation}
\label{eq:eq5.18}
s/s_0 \rightarrow\infty , \ \  \ \ t/t_0 \rightarrow\infty .
\end{equation}
The limit (\ref{eq:eq5.18}) corresponds to the increasing of the number of
$\epsilon$-separated states with the growth of time and decreasing of this
number with the growth of the separation distance. For example $\alpha =1$
and $\beta = 1/2$ correspond to the normal one-dimensional diffusion, while
other powers appear for the anomalous diffusion.

Following the definition (\ref{eq:eq5.1}) let us introduce the
$(\epsilon ;t,s)$-entropy:
\begin{equation}
\label{eq:eq5.19}
S(\epsilon ;t,s) =\ln C(\epsilon ;t,s) = -\alpha\ln (s/s_0 ) +\beta\ln
(t/t_0 ) +\ln g(\epsilon ;s,t)
\end{equation}
This expression will be analyzed in the next section.

Here we would like to mention that the complexity in the form
(\ref{eq:eq5.17}) suggest that the corresponding partition function
\begin{equation}
\label{eq:eq5.20}
{\cal N} (\epsilon ;t,s) = {\rm const.} C(\epsilon ;t,s) = {\rm const.}
(t^{\beta} /s^{\alpha} ) g(\epsilon ;t,s)
\end{equation}
has a self-similar dependence on $(t,s)$ in the limit (\ref{eq:eq5.14}),
and consequently for its moment we have
\begin{equation}
\label{eq:eq5.21}
\langle s^{\alpha} \rangle = {\cal D} \cdot t^{\beta}
\end{equation}
with $\cal D$ that may depend slowly on $t$. The equation (\ref{eq:eq5.21})
describes the anomalous diffusion with the transport exponent
\begin{equation}
\label{eq:eq5.22}
\mu = 2\beta /\alpha
\end{equation}
that can be obtained from
\begin{equation}
\label{eq:eq5.23}
(\langle s^{\alpha} \rangle )^{2/\alpha} \sim t^{\mu}
\end{equation}
(see more in~\cite{z1}).

The new parameters $\alpha ,\beta$, and $\mu$ can be considered as new
dynamical invariants intimately linked to the $(\epsilon ;t,s)$-complexity
and the corresponding entropy. Their interpretation is given in the next
section.

\subsection{Mixed complexity and the entropy extensiveness}
\label{sec:sec5.4}

Dependence of the $(\epsilon ;t,s)$-complexity on two variables $s$ and $t$
makes it possible to have ``mixed'' complexities of the following types:
\[
C^{(1)} (\epsilon ;t,s) = (s/s_0 )^{\alpha} e^{ht} g^{(1)}
(\epsilon ;t,s)
\]
\begin{equation}
\label{eq:eq5.24}
C^{(2)} (\epsilon ;t,s) =
e^{xsw} (t/t_0 )^{\beta} g^{(2)} (\epsilon ;t,s)
\end{equation}
i.e. algebraic in one variable and exponential in the second one. The
case $C^{(1)} (\epsilon ;t,s)$ shows exponential divergence in time and a
fractal length structure with a box dimension $\alpha$. The case
$C^{(2)} (\epsilon ;t,s)$ does not have yet analogy or physical
interpretation.

Similar definitions can be done for the entropy:
\[
S^{(1)} (\epsilon ;t,s) = \ln C^{(1)} (\epsilon ;t,s)
\]
\begin{equation}
\label{eq:eq5.25}
S^{(2)} (\epsilon ;t,s) = \ln C^{(2)} (\epsilon ;t,s)
\end{equation}

On the basis of all definitions of entropy we can make an important remark
about its extensivity. Typically, the extensivity is considered with
respect to the coordinate variable in statistical physics. Dynamical systems
analysis provides an extensivity with respect to the number of degrees of
freedom, or with respect to time. Here we would like to mention that instead
of $t,s$ can be $\ln t$ or $\ln s$, or different combinations. For example,
in the case of $C^{"(1)}$ the entropy $C^{(1)}$ is additive in the space of
variables $(\ln s,t)$ while in the case (\ref{eq:eq5.17}) the entropy is
additive in space $(\ln t,\ln s)$.

Our final remark for this chapter is that $(\epsilon ,t,s)$-complexity is
considered for a system with evolution and it may happen very typically,
that there are two time intervals: at the beginning the evolution of the
complexity is exponential and, after a fairly long time it becomes algebraic
[ENZ]. Similar pattern occurs for the Sinai billiard where a distribution
of the Poincar\'{e} recurrences follows exponential law for the beginning and
then decays as $\sim 1/t^3$.

\section{Traveling Waves and Riemann Invariants of Entropy and Complexity}
\label{sec:sec6}

Consider new variables
\begin{equation}
\label{eq:eq6.1}
\xi = \ln (s/s_0 ) \ , \ \ \ \ \ \eta = \ln (t/t_0 )
\end{equation}
and rewrite (\ref{eq:eq5.17}) as
\begin{equation}
\label{eq:eq6.2}
S(\epsilon t,s) =  (-\alpha\xi + \beta\eta ) g(\epsilon ,\xi ,\eta )
\end{equation}
where $g$ is the slow function of $\xi ,\eta$. Neglecting the derivatives of
$g$, we can conclude from (\ref{eq:eq6.2}) that
the entropy is constant along the traveling
wave-front
\begin{equation}
\label{eq:eq6.3}
\xi = (\beta /\alpha ) \eta \equiv c \eta
\end{equation}
where $c$ is the wave speed. Due to  (\ref{eq:eq5.22})
\begin{equation}
\label{eq:eq6.4}
\mu = 2c
\end{equation}
and the transport exponent can be interpreted as the double speed of the
traveling wave of the entropy or complexity. This property appears
because of the nontrivial coupling between time and phase space of
dynamical systems. Indeed, any kind of randomness, which is assumed for kinetics
in stochastic field, differs from the randomness of chaotic dynamics
since the randomization of trajectories occurs as a result of
nonlinearity and dynamical instability.

The structure of the entropy (\ref{eq:eq6.2})
or complexity function
(\ref{eq:eq5.17}) is
invariant, with respect to the renormalization of $\xi $ and $\eta$, namely
\begin{equation}
\label{eq:eq6.5}
s \rightarrow \lambda_s s \ , \ \ \ \  t\rightarrow \lambda_t t\ ,
\end{equation}
if we neglect variations of $g(\epsilon ;\xi ,\eta )$, and
if $\lambda_s ,\lambda_t$ satisfy the equation
\begin{equation}
\label{eq:eq6.6}
\mu /2 = c = \beta /\alpha = \ln \lambda_x /\ln \lambda_t
\end{equation}
Equation (\ref{eq:eq6.6}) shows a connection between parameters
$\alpha , \beta$ of the dynamical origin,
transport exponent $\mu$, and  velocity $c$ of the
traveling wave of the complexity. The parameter $c$ also defines a
direction in $(t,s)$ coordinates of the traveling wave propagation or,
in other words, directional complexity and entropy. For different dynamical models
scaling parameters $\lambda_s ,\lambda_t$ were defined in
[ZEN, KZ, BKWZ].

In fact, $(\alpha ,\beta )$ or $(\lambda_s ,\lambda_t )$ are not
constants, and wave propagation can be in different directions. This
situation is an analog to multi-fractal space-time structure.
For any direction with
``velocity'' $c$ in $(\xi ,\eta )$ space, there is a corresponding
curve in $(s,t)$ space and an isoline of the section of the surface
$S(t,s)$ by a plane $S(t,s)$  = const. The fractional exponents
$(\alpha ,\beta )$
of the complexity space-time dependence
are receiving a new meaning as entropy/complexity characteristics
\begin{equation}
\label{eq:eq6.7}
{\partial S \over \partial \xi} = {\partial \ln C\over \partial\xi}
= - \alpha \ , \ \ \ \ \
{\partial S \over \partial \eta} = {\partial \ln C\over \partial\eta}
=  \beta \ ,
\end{equation}
along which
the traveling waves have a constant velocity.
In this way we have a generalization of
the usual notion of entropy for a finite system.
This analogy can be advanced further by comparing waves of
complexity/entropy to  simple waves in fluid dynamics. Then
the velocity $c$ is the Riemann invariant and so is $\mu$.

We arrive to an interesting conclusion: the transport exponent $\mu$ can be
interpreted as a Riemann invariant of the complexity (entropy) simple
wave in the $(t,s)$-space. In the following chapter it will be discussed
a possibility of anisotropic phase space with different values of $\mu$ in
different directions. Correspondingly, we will have different Riemann
invariants.

\section{Directional complexity and entropy}
\label{sec:sec7}

The case (\ref{eq:eq5.17}),(\ref{eq:eq5.19}) of the self-similar behavior
of the complexity function shows explicitly that the entropy can be
written in the form
\begin{equation}
S(\epsilon ;t,s) \cong S(\xi -c\eta )
\label{eq:eq7.1}
\end{equation}
up to a slow varying function $g$. The expression (\ref{eq:eq7.1}) permits
to consider  joint space-time variables as a 2-dimensional space
$(\xi ,\eta )$ where entropy propagates along a line in the direction with
tangent $c$. As it was mentioned in the previous section, it can be
different Riemann invariants and so is for $c_{\bf e}$ where $\bf e$ is
a unit vector along some direction.

\subsection{Definitions}
\label{sec:sec7.1}

To formalize the notion of the directional complexity and entropy, let
us use an idea of Milnor
[M1,M2] for construction of directional windows in space-time (see also
[ACMF]).
Consider dynamical systems
generated by a map $f: \ I\!\!R^d \rightarrow
I\!\!R^d$, the $d$-dimensional
Euclidean space. Denote by $W$ a compact subset of $I\!\!R^d$
and by $\bf e$ a unit vector in $I\!\! R^d \times I\!\!R$ such that
the projection of $\bf e$ onto the last coordinate axis is non-negative.
Given $T\in I\!\!R$ let $W_{T,{\bf e}}$ denotes the following window
in $I\!\!R^d \times I\!\!R$:
\begin{equation}
\label{eq:eq7.2}
W_{T,{\bf e}} := \{ y\in I\!\!R^d ,
 t\in I\!\!R ,
y = x+t {\bf e} , x \in W , 0 \leq t \leq T \} \ .
\end{equation}
For example if $d = 1$, ${\bf e} = (\cos\theta ,\sin\theta )$,
$\theta \in [0,\pi ]$, then
\begin{equation}
\label{eq:eq7.3}
W_{T,{\bf e}} = \{ (x+t\cos\theta ,t\sin\theta ), 0 \leq t \leq T,
x \in W \}
\end{equation}
If $d = 2$, ${\bf e} = (\cos\alpha ,\cos\beta ,\cos\gamma )$,
$\cos^2 \alpha + \cos^2 \beta + \cos^e \gamma =1 $, then
\begin{equation}
\label{eq:eq7.4}
W_{T,{\bf e}} := \{ (x_1 + t \cos\alpha , x_2 + t\cos\beta ,
t\cos \gamma ), (x_1 ,x_2 ) \in W, 0 \leq t \leq T \} \ ,
\end{equation}

Let us generalize the Definition 2 in Section (\ref{sec:sec4.1}) of the
$\epsilon$-separability of orbits as their distinguishability through
the window $W_{T,{\bf e}}$.
\vspace{.2in}

{\bf Definition 6.} Two orbits $\{ f^n x \}$ and $\{f^n y\}$, $x,y\in W$,
are  $(\epsilon ,W_{T,{\bf e}})$-separated at time $T$ if
$(f^n x,n)\in W_{T,{\bf e}},(f^n y,n) \in W_{T,{\bf e}}$ and
dist$(f^n x,f^n y) \geq \epsilon$, where dist is the distance in
$I\!\!R^d$.

Now we may define the maximal number of
$(\epsilon ,W_{T,{\bf e}})$-separated
orbits:
\begin{equation}
\label{eq:eq7.5}
C_{\epsilon} (W_{T,{\bf e}} ) = \max \{ \# {\cal K}: \ {\cal K} \ {\rm is \ an} \
(\epsilon ,W_{T,{\bf e}})-{\rm separated \ set} \}
\end{equation}
and call it {\bf $(\epsilon ,T)$-complexity in the direction $\bf e$}.

\vspace{3ex}

Of course, the complexity depends on the set $W$ of initial points. But
the rate of its growth with $T$ may be independent of $W$. If this growth
is exponential we may define the directional topological entropy.

The topological entropy in the direction $\bf e$ is
defined by
\begin{equation}
\label{eq:eq7.6}
H_{\bf e} = \lim_{\epsilon\rightarrow 0} \sup_W
\overline{\lim}_{T\rightarrow\infty} {1\over T} \log C_{\epsilon}
(W_{T,{\bf e}} )
\end{equation}
where supremum is taken over all compact subsets $W$ of $I\!\!R^d$.

\vspace{3ex}

Note that the limit in $\epsilon$
exists since  $C_{\epsilon}
(W_{T,{\bf e}})$ is a nondecreasing function of $W$ and $\epsilon$.
Moreover $H_{\bf e} \in
[0,\infty ]$. For some systems it may happen that $H_{\bf e} = \infty$.
In this case we may compute the following quantity.

The density of topological entropy in the direction
$\bf e$ is defined by
\begin{equation}
\label{eq:eq7.7}
h_{\bf e} =
\lim_{\epsilon\rightarrow 0} \
\overline{\lim}_{{\cal D}\rightarrow \infty}
{1 \over {\cal D}^d} \left( \sup_{{\rm diam} \ W\leq {\cal D} }
\left( \overline{\lim}_{T \rightarrow\infty} {1\over T} \log C_{\epsilon}
(W_{T,{\bf e}} \right)\right)
\end{equation}

Let us remark that definitions of directional topological entropy and its
density in [ACFM] is similar to the Definitions above.

\subsection{Two Examples}
\label{sec:sec7.2}

\begin{figure}[h]
\includegraphics[width=0.5\textwidth]{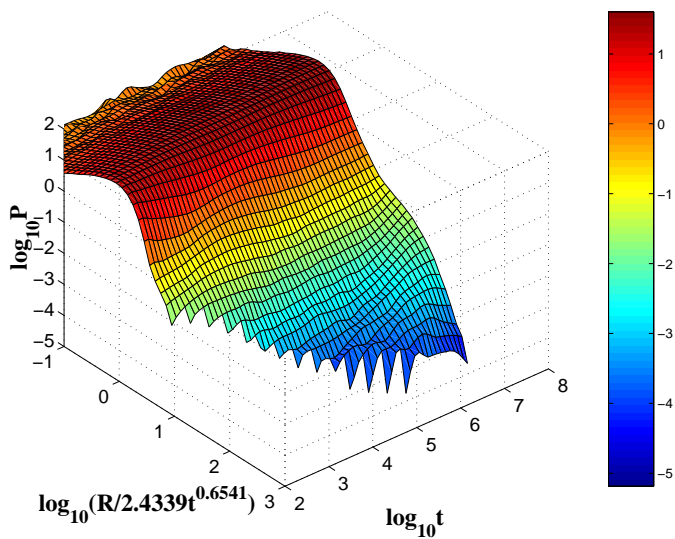}
\caption{Distribution of displacements $P = P(R,t)$ for the web map
$(R^2 = u^2 + v^2 )$ shows the complexity traveling waves.}
\label{fig3}
\end{figure}

Here we show two numerical examples that demonstrate the existence of the
directional complexity and complexity/entropy traveling waves. Both
examples are for the web map of 4th-fold symmetry [Z]. The map acts in
$(u,v)$-space:
\begin{equation}
\label{eq:eq7.8}
\bar{u} = v, \ \ \ \bar{v} = -u-K \sin v
\end{equation}
with a parameter $K$ and $u,v \in (-\infty ,\infty )$. It was shown in
[WZ] that transport exponent $\mu$ can be different in different directions of
$(u,v)$-space (see Fig.~\ref{fig3}). Dependence of transport exponents on the
the direction also appears to be recognizable for tracers in a
3-vertex flow~\cite{kz}. After we have made a link between the
$(\epsilon ;t,s)$-complexity and transport, one may expect that the
presence of the directional anisotropy indicates the
anisotropy of the complexity and
entropy as well.
More accurately we should say that for an
infinite dense set of values of $K$ the trajectories are sticky to some
islands and, due to that, the distribution function has power tails
[ZEN]. We assume that the complexity function should be of the
polynomial type for these cases. Different transport exponents for
different directions in such a case were demonstrated in [WZ] by
simulations (see Fig.~\ref{fig5}).

\begin{figure}[h]
\includegraphics[width=0.5\textwidth]{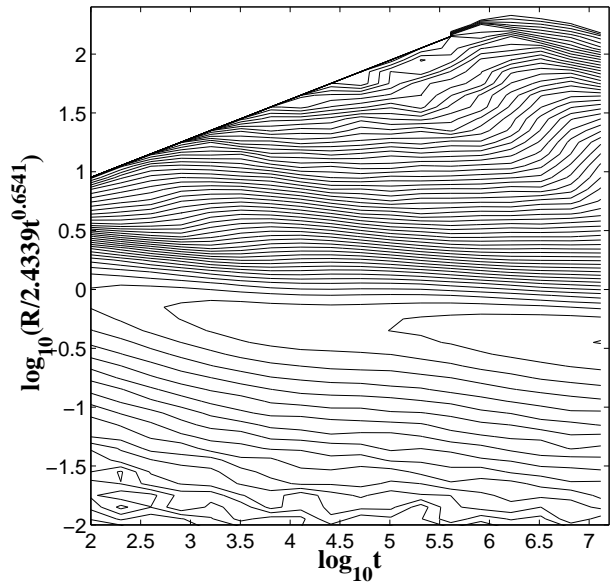}
\caption{The contour plot $P(R,t) =$ const. for the same distribution
as in Fig. 4 shows the lines of the wave propagation.}
\label{fig4}
\end{figure}

Figure~\ref{fig4} provides more information since it shows in
3-dimensional plot a
dependence of the normalized number of trajectories $P = P(R,t)$ as a
function of their position $R$ at time instant $t$ in the log-log-log
scales. The plot clearly indicates directional waves as ripples of $P$.
Figure~\ref{fig5} shows different slopes of the isolines $P$ = const
that correspond 
to different velocities of the wave propagation.

\begin{figure}[ht]
\includegraphics[width=0.5\textwidth]{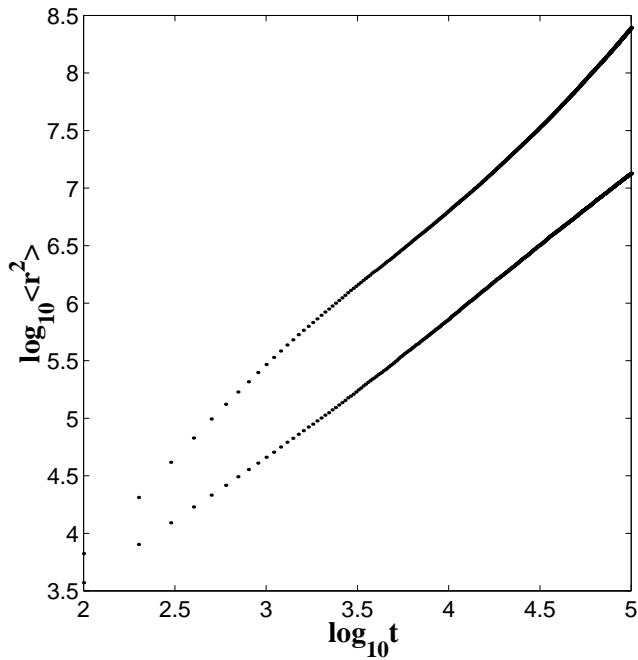}
\caption{Time dependence of the second moment for the web map shows different
slopes with transport exponent $\mu = 1.25$ along the
$u$-direction and $\mu$ = 1.50 along the diagonal direction [WZ].}
\label{fig5}
\end{figure}


\subsection{Directional complexity and rotation intervals}
\label{sec7.3}

For dynamical systems in the circle rotation numbers reflect some
properties of behavior of orbits in space-time. It is interesting to study
their relation to directional complexity. For that we consider the
map $F:I\!\!R\rightarrow I\!\!R$ of the form
\begin{equation}
F(x)=x+\omega+k\phi(x)
\end{equation}
where $\phi$ is a 1-periodic smooth function such that $|\phi'|<1$ and
$k\geq 0$ is a parameter. If $k<1$ then $F$ is one-to-one and the map
$F(x)$ mod 1 can be treated as either a continuous map of the circle
or a piecewise continuous map of the interval $[0,1]$. We choose the last
interpretation.

For $k=0$ we just have an interval exchange transformation of the interval.
For $0<k<1$ this transformation becomes nonlinear but topologically very
similar to that for $k=0$.

Consider the case $k=0$ more carefully. The orbit $x_n=F^n x_0$ for $k=0$
satisfies the equation
\begin{equation}
x_n=x_0+n\omega .
\end{equation}
Therefore the rotation number of the map $F$ equals 
$\lim_{n\rightarrow\infty} \frac{x_n}{n}=\omega$. In any window 
$W_{T,{\bf e_0}}$, ${\bf e_0}=(\cos \theta_0, \sin \theta_0)$,
$\tan \theta_0=\omega$, there are points $(x_n,n)$ for any $n>0$,
but if ${\bf e}=(\cos \theta, \sin \theta)$, $\tan \theta\neq\omega$,
then for any fixed $\epsilon>0$ there is $N>0$ such that there are
no points $(x_n,n)$, $n\geq N$, in this window. More precise, if
${\bf e}\neq{\bf e_0}$, $W=[0,1]$, then
$C_\epsilon (W_{T,{\bf e}})=C_\epsilon (W_{T_0,{\bf e}})$ for all 
$T\geq T_0=\frac{1}{|\omega-\tan \theta|}$, and the set 
$W_{T,{\bf e}}\backslash W_{T_0,{\bf e}}$ does not contain points $(x_n,n)$,
$x_0\in W$, $T_0<n\leq T$. If ${\bf e}={\bf e_0}$ then the set 
$W_{T,{\bf e}}\backslash W_{T_0,{\bf e}}$ contains points $(x_n,n)$
for any $n<T$ but the complexity $C_\epsilon (W_{T,{\bf e_0}})$ is still
bounded; in fact,  $C_\epsilon (W_{T,{\bf e_0}})\sim \frac{1}{\epsilon}$. 

The case $0<k,1$ is similar to the just described one and we omit its
consideration.

Now let $k>1$. We consider the rotation interval of the map $F$, i.e. the
set of accumulation points of sequences $\frac{F^n x}{n}$ for all
$x\in I\!\! R$, (see for instance \cite{alm}, \cite{ke}, \cite{ma1}).
This interval can be nontrivial, say $[\omega_1,\omega_2]$,
$\omega_2>\omega_1$, and for every $\omega\in [\omega_1,\omega_2]$
there
is an infinite $F$-invariant set $\Lambda_\omega$ such that 
$x\in \Lambda_\omega$ implies
$\lim_{n\rightarrow\infty} \frac{F^n x}{n}=\omega$. Moreover $\Lambda_\omega$
may contain uncountably many points (\cite{ma2}).
Trajectories in  $\Lambda_\omega$ may behave chaotically and the complexity
 $C_\epsilon (W_{T,{\bf e}})$, 
 ${\bf e}=(\cos \theta, \sin \theta)$, $\tan \theta\neq\omega$, may
grow exponentially with $T$, i.e. $H_{\bf e}>0$.

But if $\tan \theta \notin [\omega_1,\omega_2]$ then
$C_\epsilon (W_{T,{\bf e}})$ will be bounded (as above) and $H_{\bf e}$
will be 0. Thus the behavior of directional complexity has nontrivial
dependence on the direction in space-time.

The authors of \cite{sy} studied observable values of rotation
numbers. They found numerically that even if $k\gg 1$ one is able to
observe the only one (or a small number) of rotation numbers corresponding
to preferable directions in space-time.


\section{Conclusions}
\label{sec:sec89}

Nonergodicity of phase space of Hamiltonian dynamics is due to the
presence of islands. The boundaries of the islands are sticky and this imposes
strong nonuniformity of the distribution function in space and in time.
Such properties of the dynamics necessitates a new approach to the
problem of complexity, entropy and related notions of the unstable dynamics.
In this paper the notion of complexity function is introduced as a
characteristic number of mutually $\epsilon$-separated trajectories during
time $t$. This notion resembles the similar features
as a propagator function,
i.e. it depends
on the initial and final space-time. In fact, the complexity function shows
how fast trajectories disperse from each other due to the local
instability properties of a system.

The entropy can be introduced in a natural way as a logarithm of the
complexity function but the choice of variables can be different depending
on the type of local instability. In this way we were able to include
into the same scheme systems with strong intermittency and systems with
zero Lyapunov exponent.

In this article we introduced local and semi-local and flight complexity
functions that can be calculated for specific systems. The question
arises: why do not try to calculate just $\epsilon$-complexity function
and work directly with it? We think that this is impossible. The main reason
becomes clear from the following consideration.

Fix $n \gg 1$ and assume that we found an optimal set $Q_N=\{x_k\}_{k=1}^N$,
$N=C(\epsilon,n)$, of initial $(\epsilon,n)$-separated points. If we
are interested in 
the complexity function $C(\epsilon,n)$ as a function of $n$, then, say,
for $n'<n$ we either need to find a new optimal set or to use points
in $Q_N$. Of course the second option is preferable. So, the problem
appears: knowing that $Q_N$ is ($\epsilon$, n)-separated, select the maximal
subset in $Q_N$ that is ($\epsilon$, $n'$)-separated.

Let us reformulates this problem as follows. Define the graph $G$ with
vertices $(x_1,...,x_N)$ such that the edge $(x_i,x_j)$ exists if and
only if the points $x_i$ and $x_j$ are ($\epsilon$, $n'$)-separated. The
problem is to find a subgraph of $G$ with maximal number of vertices
such that every two vertices of this subgraph are connected by an edge.
This is the well-known Clique Problem which is happened to
be so called NP-complete (see for instance~\cite{gj}). Specialists
believe that 
NP-complete problems have no good algorithms, i. e.  one can not solve
such a problem by using best computers even for not so large values of
$n'$.

We see that the problem of calculation of $\epsilon$-complexity functions
is very closely related to NP-complete problems.
It is one of reasons why we introduce local and semi-local complexities.

\vspace{1.5cm}

\noindent
{\bf ACKNOWLEDGMENTS} 

We would like to thank M. Edelman for the help in preparing the figures
and M. Courbage for interesting discussions. GZ was supported by the U.S.
Navy through Grant Nos. N00014-96-1-0055 and N00014-97-1-0426, and the
U.S. Department of Energy through Grant No. DE-FG02-92ER54184. V. A. was
partially supported by CONACyT grant 485100-5-36445-E.

\end{document}